# Knowledge-aided Two-dimensional Autofocus for Spotlight SAR Filtered Backprojection Imagery

Xinhua Mao, Lan Ding, Yudong Zhang, Ronghui Zhan, and Shan Li

*Abstract*—Filtered backprojection (FBP) algorithm is a popular choice for complicated trajectory SAR image formation processing due to its inherent nonlinear motion compensation capability. However, how to efficiently autofocus the defocused FBP imagery when the motion measurement is not accurate enough is still a challenging problem. In this paper, a new interpretation of the FBP derivation is presented from the Fourier transform point of view. Based on this new viewpoint, the property of the residual 2-D phase error in FBP imagery is analyzed in detail. Then, by incorporating the derived *a priori* knowledge on the 2-D phase error, an accurate and efficient 2-D autofocus approach is proposed. The new approach performs the parameter estimation in a dimension-reduced parameter subspace by exploiting the *a priori* analytical structure of the 2-D phase error, therefore possesses much higher accuracy and efficiency than conventional blind methods. Finally, experimental results clearly demonstrate the effectiveness and robustness of the proposed method.

## I. Introduction

SYNTHETIC aperture radar (SAR) obtains high spatial resolution by exploiting the motion of a small real antenna to synthesize an equivalent larger aperture antenna. This aperture synthesis is implemented by processing coherently the raw data collected by a moving radar during the synthetic aperture time [1], [2]. When the radar platform flies along a linear trajectory and the radar collects data at constant pulse repeat frequency (PRF), the coherent processing can be performed efficiently by a batch processing using frequency domain image formation algorithm, such as range Doppler algorithm, chirp scaling algorithm and range migration algorithm. This linear flight-path assumption is often valid when the synthetic aperture time is not very long. However, as the resolution becomes finer and finer, the required synthetic aperture length becomes very long [3]-[5], or when the radar is equipped on a maneuverable platform, such as multi-rotors mini UAV [6], [7], non-linear radar flight path will become commonplace. If a frequency domain algorithm is still used in these cases, complicated motion compensation processes are required to correct for the phase error caused by the non-ideal radar motion. As an alternative, time domain correlation-based algorithms, e.g., filtered backprojection (FBP) algorithm, become more and more popular in these cases because of their inherent non-linear motion compensation capability [8]-[10].

Nevertheless, an accurate image formation processing in FBP also requires an accurate measurement of the geometric relationship between the radar's flight path and the scene being imaged. Modern SAR sensor accomplishes the measurement of radar motion using a motion sensing system consisting of some combination of an inertial measurement unit (IMU) and a global positional system (GPS) navigator [11]. These sensors, however, may be too expensive or cannot provide the satisfactory measurement accuracy for very high resolution SAR imaging. Consequently, signal based motion compensation, i.e., autofocus, is often an indispensable process in SAR processing [12]-[14].

In the literature, most efficient autofocus approaches are image post-processing techniques. They refocus the defocused imagery produced by image formation algorithm by estimating and correcting for the residual phase error in the image spectrum domain. For frequency domain image formation algorithms, the spectral characteristics of the produced image are explicit and have been thoroughly examined in the literature. Therefore, there are many well-developed autofocus algorithms, e.g., Mapdrift [15], Phase Difference [16], MCA [17], Phase Gradient Autofocus (PGA) [18], to refocus the defocused imagery produced by a frequency domain algorithm. However, for time domain image formation algorithms, due to their unclear spectral characteristics of the produced imagery, it is still a challenging problem to use efficient post-processing based autofocus approaches, e.g., the most popular autofocus method PGA, to refocus the defocused imagery. Due to this reason, most efforts in the literature are focused on the optimum theory based approaches which constantly adjust the radar's motion parameters during the image formation processing to maximize the image quality index, such as sharpness, contrast, or entropy [19]-[23]. In these approaches, the autofocus is incorporated into the image formation process. Therefore to search for the optimum motion parameters, the image formation process has to be repeated again and again. This

This work was supported in part by the National Natural Science Foundation of China under Grant 61671240, in part by Natural Science Foundation of Jiangsu Province under Grant BK20170091.

Xinhua Mao and Lan Ding are with the College of Electronic and Information Engineering, Nanjing University of Aeronautics and Astronautics, Nanjing 211106, China. (e-mail: xinhua@nuaa.edu.cn).

Yudong Zhang is with the Department of Informatics, University of Leicester, Leicester, UK (e-mail: yudongzhang@ieee.org).

Ronghui Zhan is with the College of Electronic Science and Engineering, National University of Defense Technology. (e-mail: zhanrh@nudt.edu.cn).

exhaustive search makes these optimization based algorithms possess a poor computational efficiency.

To use the high-efficient and widely-used PGA algorithm on FBP imagery, Jakowats et al. first show that there exists an approximate Fourier transform relationship between the FBP image domain and the range-compressed phase history domain when the image is formed on a range-bearing grid [24]. Recently, Doerry et al. present a detailed analysis on the basics of the backprojection algorithm [10]. In their report, the spectrum characteristics of the FBP imagery are discussed and some preprocessing steps on the FBP imagery are suggested to facilitate the autofocus processing.

Nevertheless, all these approaches can only deal with the one-dimensional defocus problem. They all assume that the residual range cell migration after image formation algorithm processing can be neglected and the autofocus only needs to estimate and correct for the 1-D azimuth phase error. This assumption becomes invalid as the resolution increase, especially when high accuracy motion sensors can't be available [3], [6], [25]. In these situations, 2-D autofocus becomes a necessary procedure to obtain refocused images. In the literature, the existing 2-D autofocus approaches can be generally divided into two categories. One is to estimate the 2-D phase error in a blind manner. They assume that the 2-D phase error is completely unknown and estimate all the 2-D phase error parameters directly [26]-[31]. Because of the large number of unknown phase parameters, these strategies often suffer from inferior computational efficiency and parameter estimate accuracy. In contrast to the blind estimation approaches, in the second strategy, the 2-D phase error is estimated in a semi-blind manner [3], [34-36]. By incorporating the *a priori* knowledge on the phase error structures, these approaches estimate the 2-D phase error in a dimension-reduced parameter space. Compared with the blind methods, these approaches possess significantly improved parameter estimation performance in both computational efficiency and estimation accuracy due to the reduced dimension of phase parameters.

For these dimension-reduced autofocus strategies, the *a priori* structure information on the phase error is a prerequisite condition. For frequency domain image formation algorithms, the spectrum characteristics of the 2-D phase error have been well investigated in recent years [3], [37], [38]. However, for time domain algorithm, although Doerry's report [10] gave a good start on the spectrum analysis for the FBP imagery, more details about the 2-D phase error after FBP processing, e.g., the ambiguity and structural property of the residual 2-D phase error, are still unknown in the literature. Therefore, how to efficiently refocus the 2-D defocused FBP imagery is still a challenging problem.

In this paper, a new interpretation of the FBP formulation is presented from the Fourier transform point of view. Based on this new viewpoint, the properties of the residual 2-D phase error in FBP imagery, including spectrum ambiguity, space-variant spectrum support and analytical structure of the 2-D phase error, are analyzed in detail. Then, by incorporating the *a priori* knowledge on the 2-D phase error, an accurate and efficient 2-D autofocus approach is proposed. The new approach performs the phase error parameter estimation in a dimension-reduced subspace by exploiting the *a priori* analytical structure of the 2-D phase error, therefore possesses much higher accuracy and efficiency than conventional blind methods.

The rest of the paper is organized as follows. In Section II, a new formulation of FBP is presented from the Fourier viewpoint. By using this new interpretation, the spectral characteristics of the FBP imagery are detailed in Section III. By exploiting the derived *a priori* information on the spectral characteristics, an efficient one-dimensional estimation/two-dimensional correction autofocus approach is proposed in Section IV. Finally, in Section V, experimental results are presented to demonstrate the effectiveness of the proposed autofocus approach. Section VI presents concluding remarks.

## II. NEW FORMULATION OF FILTERED BACKPROJECTION ALGORITHM

In this section, we present a new formulation of the filtered backprojection algorithm, which shows a clear Fourier transform relationship between the phase history domain and FBP image domain. This new interpretation is very beneficial when investigating the property of residual 2-D phase error in FBP imagery in the next section.

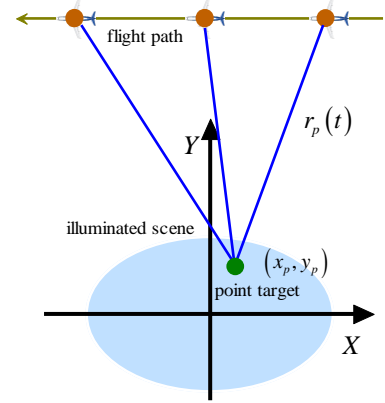

Fig.1. Radar data collection geometry.

### A. Signal model

The imaging geometry of a spotlight-mode SAR system is shown in Fig.1. The radar moving at a nominal velocity of $v$ transmits a train of coherent wideband pulses to illuminate an imaging scene. Without loss of generality, a generic point target located at $(x_p, y_p)$ in the illuminated scene is assumed, which means that the scene reflection function can be denoted as $f_0(x, y) = \delta(x - x_p, y - y_p)$. Let $t$ and $\tau$ represent the slow time and fast time, respectively. The instantaneous position of the antenna phase center (APC) of the radar is denoted as $[x_a(t), y_a(t)]$.

From the geometry, the instantaneous range from the point target to the APC can be expressed as

$$r_p(t) = \sqrt{[x_a(t) - x_p]^2 + [y_a(t) - y_p]^2}. \quad (1)$$

If we assume that the transmitted signal is a linear frequency modulation (LFM) signal modulated at a carrier frequency $f_c$, then the echo signal after demodulation can be expressed as

$$s(t,\tau) = A \cdot \text{rect}\left(\frac{\tau - 2r_p(t)/c}{T_r}\right) \cdot \exp\left\{j\pi k\left(\tau - \frac{2r_p(t)}{c}\right)^2\right\} \cdot \exp\left\{-j2\pi f_c \frac{2r_p(t)}{c}\right\}, \quad (2)$$

where $c$ is the propagation speed of the electromagnetic wave, $k$ and $T_r$ are the chirp rate and chirp width of the transmitted LFM signal, respectively.

After pulse compression in the range direction, the 2-D echo signal can be simplified as

$$s(t,\tau) = \text{sinc}\left[B\left(\tau - \frac{2r_p(t)}{c}\right)\right] \cdot \exp\left\{-j2\pi f_c \frac{2r_p(t)}{c}\right\}, \quad (3)$$

where $B = kT_r$ is the bandwidth of the transmitted signal. It should be noted that in (3) and also in the following derivation, we ignore the nonessential amplitude effect to simplify the notation.

### B. Filtered Backprojection Algorithm

Backprojection algorithm processing can be viewed as a beamforming process. First, an imaging grid which covers the scene of interest is defined, as shown in Fig.2. Then, for each pixel in the imaging grid, compute its contribution in each range-compressed pulse (by a range computation and interpolation), and then coherently accumulate them.

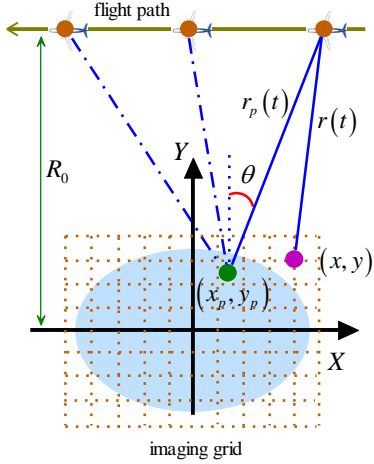

Fig.2. FBP imaging geometry.

Specifically, if the coordinate of the pixel in imaging grid is assumed as $(x,y)$, then the instantaneous range from this pixel to the APC can be computed as

$$r(t) = \sqrt{\left[x_a(t) - x\right]^2 + \left[y_a(t) - y\right]^2}. \quad (4)$$

With this range, the contribution of this pixel (if there are scatterers in this pixel) on the range-compressed pulse can be found with $s(t, 2r(t)/c)$. To coherently accumulate the contributions from all the pulses, a Doppler phase alignment should be performed firstly. Therefore, the produced backprojection imagery can be expressed as

$$f(x,y) = \int_{-T/2}^{T/2} s\left(t, \frac{2r(t)}{c}\right) \cdot \exp\left\{j2\pi f_c \frac{2r(t)}{c}\right\} dt, \quad (5)$$

where $T$ is the synthetic aperture time.

By inserting (3) into (5), we obtain

$$f(x,y) = \int_{-T/2}^{T/2} \text{sinc}\left[B\left(\frac{2r(t)}{c} - \frac{2r_p(t)}{c}\right)\right] \cdot \exp\left\{j\frac{4\pi}{c}f_c\left[r(t) - r_p(t)\right]\right\} dt. \quad (6)$$

Using Fourier transform relationship, the sinc function term in (6) can be expressed as following

$$\text{sinc}\left[B\left(\frac{2r(t)}{c} - \frac{2r_p(t)}{c}\right)\right] = \int_{-B/2}^{B/2} \exp\left\{j2\pi f_\tau\left[\frac{2r(t)}{c} - \frac{2r_p(t)}{c}\right]\right\} df_\tau, \quad (7)$$

where $f_\tau$ is the range frequency. Therefore, inserting (7) into (6), (6) can be rewritten as

$$f(x,y) = \int_{-T/2}^{T/2} \int_{k_{rc}-\Delta k_r/2}^{k_{rc}+\Delta k_r/2} \exp\left\{jk_r\left[r(t) - r_p(t)\right]\right\} dk_r dt, \quad (8)$$

where $k_r = \frac{4\pi}{c}(f_c + f_\tau)$, $k_{rc} = \frac{4\pi}{c}f_c$, and $\Delta k_r = \frac{4\pi}{c}B$.

In (8), both $k_r$ and $t$ are denoted as continuous variables. In the actual situation, we only get discrete samples in both $k_r$ and $t$ domain. These samples are often uniformly-spaced in both domains. However, after mapping into wavenumber domain, the sampling density is different for different areas in wavenumber domain. To compensate for these density variations effect, a ramp filter is included. Therefore, the resulted filtered backprojection (FBP) imagery can be expressed as

$$f(x,y) = \int_{-T/2}^{T/2} \int_{k_{rc}-\Delta K_r/2}^{k_{rc}+\Delta K_r/2} \exp\left\{jk_r\left[r(t) - r_p(t)\right]\right\} k_r dk_r dt. \quad (9)$$

### C. New Formulation of Filtered Backprojection Algorithm

From the geometry shown in Fig.2, the differential range in (9) can be approximated as

$$r(t) - r_p(t) \approx (x_p - x)\sin\theta + (y_p - y)\cos\theta, \quad (10)$$

where

$$\theta = \text{atan}\left(\frac{x_a(t) - x_p}{y_a(t) - y_p}\right). \quad (11)$$

By substituting (10) into (9), the filtered backprojection imagery can be approximated as

$$f(x,y) = \int_{-T/2}^{T/2} \int_{k_{rc}-\Delta k_r/2}^{k_{rc}+\Delta k_r/2} \exp\left\{jk_r\left[(x_p - x)\sin\theta + (y_p - y)\cos\theta\right]\right\} k_r dk_r dt. \quad (12)$$

From the definition of $\theta$ in (11) and the geometry in Fig.2, it is clear that there is a one-to-one corresponding relationship between the angle $\theta$ and the azimuth time $t$. Therefore, we can get an inverse function relationship $t = g(\theta)$ (a specific expression can be derived from (11), but is not required in the following derivation). From this relationship, we can obtain the relationship between $dt$ and $d\theta$ as following

$$dt = g'(\theta) d\theta. \quad (13)$$

Inserting (13) into (12), (12) can be rewritten as

$$f(x,y) = \int_{\theta_{start}}^{\theta_{end}} \int_{k_{rc}-\Delta k_r/2}^{k_{rc}+\Delta k_r/2} \exp\left\{j\left[(x_p - x)k_r\sin\theta + (y_p - y)k_r\cos\theta\right]\right\} k_r dk_r d\theta, \quad (14)$$

where the slow-variant amplitude factor $g'(\theta)$ can be approximated as a constant and eliminated for notation simplification.

By converting the polar coordinate $(k_r, \theta)$ into the Cartesian coordinate $(k_x, k_y)$ using the relationship $k_x = k_r \sin\theta$ and $k_y = k_r \cos\theta$, (14) can be rewritten as

$$f(x,y) = \iint_D \exp\{j[(x_p - x)k_x + (y_p - y)k_y]\} dk_x dk_y, \quad (15)$$

where $D$ is the 2-D integral interval determined by $\sqrt{k_x^2 + k_y^2} \in [k_{rc} - \Delta k_r/2, k_{rc} + \Delta k_r/2]$, $\operatorname{atan}(k_x/k_y) \in [\theta_{start}, \theta_{end}]$.

From (15), we can get the spectrum of the FBP imagery as

$$\mathbb{F}[f(x,y)] = \exp\{j[x_p k_x + y_p k_y]\}. \quad (16)$$

This is exactly the spectrum of the target function $f_0(x,y)$.

From the previous analysis, we have known that the two frequency variables in the spectrum domain are expressed as

$$\begin{aligned} k_x &= k_r \sin\theta \\ k_y &= k_r \cos\theta \end{aligned}, \quad (17)$$

where

$$\begin{aligned} k_r &= \frac{4\pi}{c}(f_c + f_\tau) \\ \theta &= \operatorname{atan}\left(\frac{x_a(t) - x_p}{y_a(t) - y_p}\right) \end{aligned}. \quad (18)$$

From (18), it is clear that the spectrum support area can be completely determined by the parameters of the transmitted signal and the relative geometry relationship between the radar and the target.

If the far-field assumption is satisfied, i.e., the radar wavefront can be approximated as planar, then the target position dependence of the angle $\theta$ can be ignored, i.e., (18) can be approximated as

$$\begin{aligned} k_r &= \frac{4\pi}{c}(f_c + f_\tau) \\ \theta &= \operatorname{atan}\left(\frac{x_a(t)}{y_a(t)}\right) \end{aligned}. \quad (19)$$

In this case, all the scatterers in the illuminated scene approximately share the same spectrum support areas, just as Fig.3(a) shown.

If the far-field assumption is not satisfied, to show the space-variant property of the spectrum support, we can perform a Taylor expansion approximation on (18) with respect to $x_p$ and $y_p$

$$\theta \approx \operatorname{atan}\left(\frac{x_a(t)}{y_a(t)}\right) - \frac{1}{\left(1 + \left(\frac{x_a(t)}{y_a(t)}\right)^2\right) y_a(t)} x_p + \frac{x_a(t)}{\left(1 + \left(\frac{x_a(t)}{y_a(t)}\right)^2\right) y_a^2(t)} y_p. \quad (20)$$

In most cases, $x_a(t) \ll y_a(t)$, $\frac{x_a(t)}{y_a^2(t)} y_p \ll \frac{1}{y_a(t)} x_p$, $\frac{1}{y_a(t)} x_p \approx \frac{1}{y_a(0)} x_p$, then (20) can also be approximated as

$$\theta \approx \operatorname{atan}\left(\frac{x_a(t)}{y_a(t)}\right) - \frac{1}{y_a(0)} x_p. \quad (21)$$

From (21), it is clear that the position of the spectrum support region is target-position dependent. Specifically, the polar angle of the spectrum support varies linearly depend on the target's azimuth coordinate. Therefore, different targets possess different spectrum supports, as shown in Fig.3(b). These skewed spectrum supports will make the autofocus process challengeable.

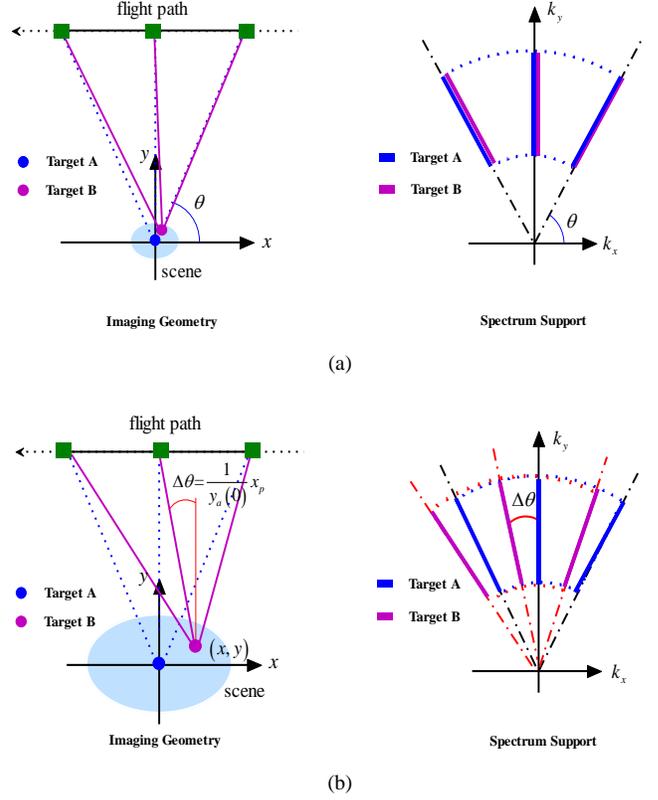

Fig.3. Data collection geometry and its corresponding spectrum support. (a) Far Field. (b) Near Field.

From the above analysis, we know that the mechanics of the FBP is very similar to the PFA. It also includes an underlying polar-to-Cartesian coordinate transformation in spatial frequency domain and then an underlying Fourier transform which transform the data from spatial frequency domain to image domain. But they also have some differences. Firstly, in PFA, the sample coordinates in spatial frequency domain for all point targets are all determined by the relative geometric relationship between the radar platform and the scene center, therefore the polar-to-Cartesian transformation is space-invariant. However, in FBP, the sample coordinate in spatial frequency domain for each point target is determined by the relative geometric relationship between the radar platform and the point target itself; therefore, the polar-to-Cartesian transformations are different for different targets. Secondly, the implementation of the Fourier transform from spatial frequency to image domain are different for the two algorithms. In PFA, Fourier transform is often implemented by FFT, while in FBP, it is implemented by a phase alignment followed by a summation. These differences will complicate the application of autofocus on FBP imagery. We will give a detail discussion in the following section.

## III. Prior Knowledge on Phase Error in FBP Imagery

Using the above new interpretation of FBP derivation, we will investigate the characteristics of the 2-D phase error in FBP imagery in this section. The derived prior information on the 2-D phase error will provide a prerequisite for the proposed 2-D autofocus approach in the next section.

### A. 2-D Phase Error in Image Spectrum Domain

In the above analysis, we assumed that the radar's positions during the synthetic aperture time are perfectly known. However, for an actual SAR system, the motion measurement subsystem, such as IMU and/or GPS, often cannot measure the radar's motion accurately. Therefore, some residual motion error remains. For an arbitrary pixel in the imaging grid, we assume that the actual range and the measured range from the APC to this pixel are $r(t)$ and $r(t)+r_e(t)$, respectively. Therefore, $r_e(t)$ is the range measurement error. During the image formation process, the measured range, instead of the actual range, is exploited. Therefore, the resulted backprojection imagery becomes

$$f(x,y) = \int_{-T/2}^{T/2} \int_{k_{rc}-\Delta k_r/2}^{k_{rc}+\Delta k_r/2} \exp\left\{jk_r\left[r(t)+r_e(t)-r_p(t)\right]\right\}k_r dk_r dt. \quad (22)$$

Compared with (9), we can see that the 2-D phase error in phase history domain can be expressed as

$$\Phi_e(t,k_r) = k_r r_e(t). \quad (23)$$

By inserting (10) into (22), (22) can be also expressed as

$$f(x,y) = \int_{-T/2}^{T/2} \int_{k_{rc}-\Delta k_r/2}^{k_{rc}+\Delta k_r/2} \exp\left\{jk_r\left[(x-x_p)\sin\theta + (y-y_p)\cos\theta + r_e(t)\right]\right\}k_r dk_r dt. \quad (24)$$

After a change-of-variable $t = g(\theta)$, (24) can be rewritten as

$$f(x,y) = \int_{\theta_{start}}^{\theta_{end}} \int_{k_{rc}-\Delta k_r/2}^{k_{rc}+\Delta k_r/2} \exp\left\{j\left[k_r\left[(x-x_p)\sin\theta \right.\right.\right.\\\left.\left.\left.+(y-y_p)\cos\theta + r_e(g(\theta))\right]\right]\right\}k_r dk_r d\theta \quad (25)$$

As before, the slow-variant amplitude factor $g'(\theta)$ is eliminated for notation simplification.

If we define a composite function $\xi(\theta) = r_e(g(\theta))$, then (25) can also be expressed as

$$f(x,y) = \int_{\theta_{start}}^{\theta_{end}} \int_{k_{rc}-\Delta k_r/2}^{k_{rc}+\Delta k_r/2} \exp\left\{j\left[k_r\left[(x-x_p)\sin\theta \right.\right.\right.\\\left.\left.\left.+(y-y_p)\cos\theta + \xi(\theta)\right]\right]\right\}k_r dk_r d\theta \quad (26)$$

Converting the polar coordinate $(k_r,\theta)$ into Cartesian coordinate $(k_x,k_y)$ using the relationship: $k_r = \sqrt{k_x^2+k_y^2}$, $\theta = \text{atan}(k_x/k_y)$, we get

$$f(x,y) = \iint_D \exp\left\{j\left[(x_p-x)k_x + (y_p-y)k_y \right.\right.\\\left.\left.+\sqrt{k_x^2+k_y^2}\,\xi\left[\text{atan}\left(\frac{k_x}{k_y}\right)\right]\right]\right\}dk_x dk_y \quad (27)$$

From (27), we can get the spectrum of the actual FBP imagery

$$\mathbb{F}[f(x,y)] = \exp\left\{j\left[x_p k_x + y_p k_y + k_y \zeta\left(\frac{k_x}{k_y}\right)\right]\right\}, \quad (28)$$

where $\zeta(\cdot)$ is a composite function defined by $\zeta(u) = \sqrt{1+u^2}\,\xi[\text{atan}(u)]$.

Comparing (28) with (16), we can get the 2-D phase error in the spectrum domain for the actual FBP imagery

$$\Phi_e(k_x,k_y) = k_y \zeta\left(\frac{k_x}{k_y}\right). \quad (29)$$

### B. Ambiguity Property of 2-D Phase Error

From the above subsection, we know that the 2-D phase error in spatial frequency domain is $\Phi_e(k_x,k_y)$, whose expression is presented in (29). It should be noted that the range spatial frequency $k_y$ has a constant offset $k_{yc}$, i.e., $k_y = k_{yc} + \bar{k}_y$, where $\bar{k}_y \in \left[-\Delta k_y/2,\ \Delta k_y/2\right]$.

To correct for the 2-D phase error, we have to return to the spatial frequency domain from the FBP image domain. This transformation is often implemented by FFT. It should be noted that the spatial frequency variable in FFT has no offset. That is to say, the actual spectrum has an offset in the range frequency domain, but the observed spectrum domain by FFT is limited to the baseband. Therefore, an ambiguity will exist because the offset frequency $k_{yc}$ often far exceeds the sample rate $k_{ys}$. This ambiguity effect can be clearly illustrated in Fig.4.

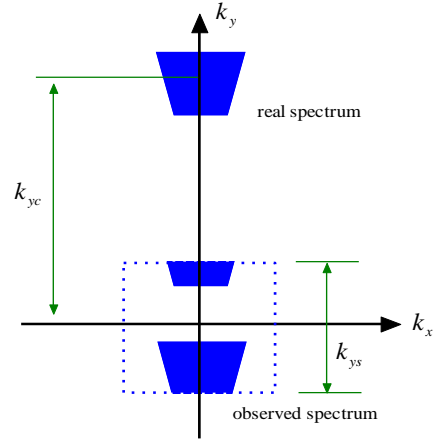

Fig.4. Illustration of sample spectrum ambiguity in range spatial frequency domain.

Why doesn't this ambiguity effect happen in PFA and RMA? In PFA and RMA, the transformation from spatial frequency domain to image domain is implemented by IFFT. Therefore an FFT on image data can reconstruct the spatial spectrum unambiguously because FFT is an inverse operation of IFFT. However, in FBP, we know in the previous section that the transformation from spatial frequency domain to image domain is implemented by a phase alignment and summation. It is also an inverse Fourier transform, but different from the IFFT, the spectrum range in this implementation includes the actual spectrum support. So in this situation, an FFT on the FBP imagery can't reconstruct the spatial spectrum unambiguously because FFT is not the accurate inverse operation of this inverse Fourier transform. The difference between the two algorithms can be clearly illustrated by Fig. 5.

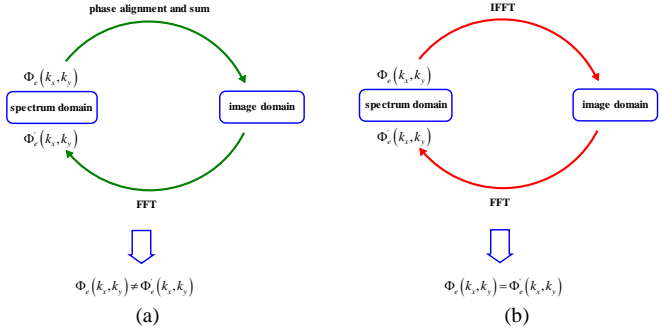

Fig.5. Relationship between spectrum domain and image domain in different image formation algorithms. (a) FBP. (b) PFA.

*C. Space-variant Property of 2-D Phase Error*

In the literature, most of efficient autofocus algorithms all assume that the phase errors are space-invariant, i.e., the phase errors are assumed common to all targets. However, this assumption is hard to meet for the FBP imagery.

In phase history domain, the 2-D phase error is linearly related to the range error between the radar and the scatterers, i.e.,

$$\Phi_e(t,k_r) = k_r r_e(t), \quad (30)$$

where

$$r_e(t) = \sqrt{\left[x_a(t) + x_e(t) - x_p\right]^2 + \left[y_a(t) + y_e(t) - y_p\right]^2} - \sqrt{\left[x_a(t) - x_p\right]^2 + \left[y_a(t) - y_p\right]^2}. \quad (31)$$

Strictly speaking, this phase error is space-variant, that is, the phase errors are different for different scatterers. Fortunately, this space-variant effect is usually small and generally not of consequence for most actual applications. This is also the underlying assumption for almost all the existing autofocus algorithms.

However, for autofocus algorithms which operated as a post-processing process, such as PGA, we are often more concerned with the phase error in the image spectrum domain instead of the phase history domain. From the previous section, we have known that the mapping from phase history domain to spatial frequency domain in FBP can be divided into two processes, i.e.,

$$(t,k_r) \xrightarrow{\theta \approx \mathrm{atan}\left(\frac{x_a(t)}{y_a(t)}\right) - \frac{1}{y_a(0)}x_p} (\theta,k_r) \xrightarrow{k_x = k_r \sin\theta,\; k_y = k_r \cos\theta} (k_x,k_y). \quad (32)$$

It first maps the data from phase history domain $(t,k_r)$ to polar format spatial frequency domain $(\theta,k_r)$ by a target-position-dependent transformation $\theta \approx \mathrm{atan}(x_a(t)/y_a(t)) - x_p/y_a(0)$, then it maps the data from polar to Cartesian by a target-position-independent transformation: $k_x = k_r \sin\theta$; $k_y = k_r \cos\theta$.

We assume that there are two point targets (A and B) located in the scene, target A is located in the scene center whose coordinate is $(0,0)$, and target B is located at $(x_p,y_p)$. Their 2-D phase error in phase history domain are denoted as $\Phi_e^A(t,k_r)$ and $\Phi_e^B(t,k_r)$, respectively. We also assume that the phase errors in phase history domain are approximately space-invariant. Therefore, the relationship of the 2-D phase error in phase history domain between the two targets are $\Phi_e^B(t,k_r) = \Phi_e^A(t,k_r)$. That is, the 2-D phase errors for different targets are the same in the phase history domain.

However, after mapping from phase history domain to spatial frequency domain as shown in (32), it is easy to get that the analytical relationship of the 2-D phase error between the two targets will become

$$\Phi_e^B(k_x,k_y) = \Phi_e^A(k_x \cos\Delta\theta + k_y \sin\Delta\theta, -k_x \sin\Delta\theta + k_y \cos\Delta\theta), \quad (33)$$

where $\Delta\theta = x_p/y_a(0)$. This mapping process can also be graphically illustrated in Fig.6.

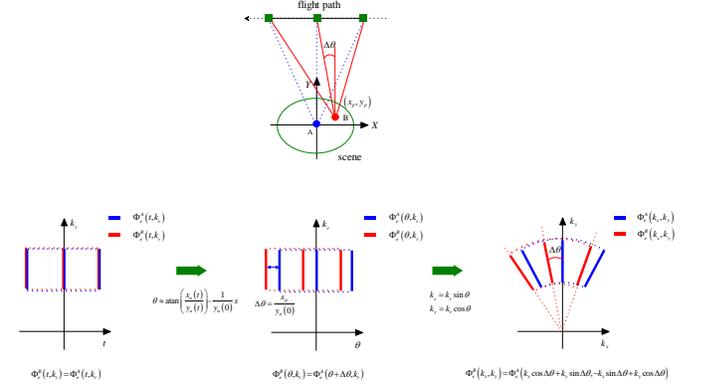

Fig.6. Support area of 2-D phase error in different domains.

In many actual imaging situations, the illuminated scene size is often much smaller than the standoff range, then $\Delta\theta = x_p/y_a(0)$ is very small, therefore, $\cos\Delta\theta \approx 1, \sin\Delta\theta \approx \Delta\theta$. Using these approximations, (33) can be rewritten as

$$\Phi_e^B(k_x,k_y) \approx \Phi_e^A(k_x + k_y \Delta\theta, k_y). \quad (34)$$

*D. Structure Property of 2-D Phase Error*

From (29), we can see that the residual phase error in FBP imagery is essentially two-dimensional. In most cases, this 2-D phase error can be approximated as a one-dimensional azimuth phase error. That is to say, the effect of range-frequency variation on the 2-D phase error can be ignored. However, when the resolution becomes very fine, and/or the motion sensor has poor measurement accuracy, this approximation is often invalid. To get an accurately refocused imagery, it is necessary to estimate and correct for the whole 2-D phase error. To show clearly the effect of the 2-D phase error on the focus property of the target, we can perform a Taylor expansion on (29) with respect to the range frequency evaluated at the center of the range frequency $k_{yc} = 4\pi f_c / c$:

$$\Phi_e(k_x,k_y) = \phi_0(k_x) + \phi_1(k_x)(k_y - k_{yc}) + \phi_2(k_x)(k_y - k_{yc})^2 + \cdots, \quad (35)$$

where $\phi_0(k_x) = k_{yc}\zeta(k_x/k_{yc})$ represents the azimuth phase error, it causes target defocus in the azimuth direction; $\phi_1(k_x) = \zeta(k_x/k_{yc}) - k_x/k_{yc}\zeta'(k_x/k_{yc})$ is the residual RCM, if not compensated, it causes 2-D defocus; $\phi_2(k_x) = k_x^2/(2k_{yc}^3)\zeta''(k_x/k_{yc})$ and other high-order terms are related to the range defocus.

If we have no any prior knowledge on the 2-D phase error, we have to estimate the whole 2-D phase error directly, or equivalently estimate the APE, residual RCM, and the range defocus terms as an approximation. Due to high parameter dimension, these approaches often suffer from high computational complexity and poor parameter estimate accuracy in actual application. Fortunately, from (29), we can see that the 2-D phase error is not absolutely unknown. In fact, in (29) only the 1-D function $\zeta(u)$ is unknown. This unknown function is linearly related with the APE as $\phi_0(k_x) = k_{yc}\zeta(k_x/k_{yc})$. Therefore, we can also express the 2-D phase error as a function of the APE

$$\Phi_e(k_x,k_y) = \frac{k_y}{k_{yc}}\phi_0\left(\frac{k_{yc}}{k_y}k_x\right). \quad (36)$$

This equation shows that to estimate the 2-D phase error we need only to estimate the 1-D APE directly.

## IV. KNOWLEDGE-AIDED TWO-DIMENSIONAL AUTOFOCUS FOR FBP IMAGERY

To get an accurately focused image by post-processing the defocused FBP imagery, it is necessary to estimate and correct for the 2-D phase error $\Phi_e(k_x,k_y)$ in image spectrum domain. When the 2-D phase error is not very large, it may be approximated as a 1-D azimuth phase error. In this case, a traditional one-dimensional autofocus processing on the defocused imagery will be accurate enough to get a well-focused imagery. In this paper, however, we will take into account the general case, that is, the 2-D phase error must be all corrected for.

To estimate the 2-D phase error, one possible approach is to estimate all the 2-D phase error parameters directly. However, this kind of blind estimation approach often suffers from low estimate accuracy and high computational complexity due to the high dimensionality of the unknown parameters. Fortunately, from the previous section, we have learned some specific property about the 2-D phase error. If we can exploit this *a priori* knowledge, the estimation process can be greatly simplified. For example, if we exploit the structure property of the 2-D phase error and eliminate the spectrum ambiguity, the 2-D phase error estimation can be reduced into a 1-D phase error estimation. This will greatly reduce the computational complexity and improve the estimate accuracy. Also, if we exploit the *a priori* knowledge on space-variant property of the 2-D phase error and perform a preprocessing to eliminate this space-variation, the phase error estimate and correction can be implemented by a high-efficient batch processing. Due to these reasons, we propose an efficient semi-blind 2-D autofocus approach. The new approach includes four main processes. Firstly, two preprocessing operations including spectrum ambiguity elimination and spectrum alignment are performed to facilitate the following phase error estimate and correction. Then, by incorporating the *a priori* phase structure information, the 2-D phase error estimation is implemented efficiently in a low dimensional subspace. That is, only 1-D azimuth phase error is estimated directly from the image data, while the 2-D phase error is then computed from this estimated APE by exploiting the phase structure information. The whole flowchart of the new approach is shown in Fig.7.

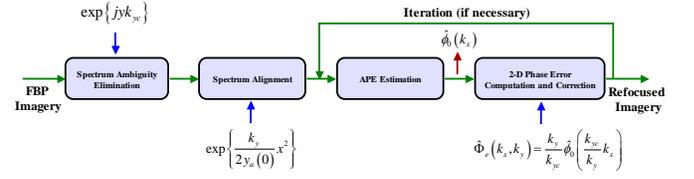

Fig.7. Flowchart of the proposed knowledge-aided 2-D autofocus approach.

### A. Spectrum Ambiguity Elimination

For a 2-D autofocus approach whose input is a defocused FBP imagery, it is necessary to return to the spectrum domain because the 2-D phase error estimate and correction are all performed in this domain. This transformation is often implemented by FFT due to its high computational efficiency.

From section III.C, we have known that the FBP image spectrum has an offset in the $K_y$ domain and the transformation from spectrum domain to image domain in the FBP derivation has accounted this offset effect. So if we return to the spectrum domain from the FBP image domain by a FFT, the spectrum will be aliased into the baseband because FFT doesn't take into account the spectrum offset.

This spectrum ambiguity can be disregarded in 1-D azimuthal autofocus, because a constant ambiguity in range dimension will not affect the estimation and correction for the azimuth phase error. However, in our proposed 2-D autofocus approach, this ambiguity must be addressed because a range frequency dependent mapping is included in the 2-D autofocus process.

To eliminate the spectrum ambiguity, a phase correction should be performed in the image domain to down-convert the data into baseband. Because the image spectrum is only offset in range frequency by $K_{yc}$, then the phase correction function is

$$f_{cor1}(x,y) = \exp\{jyk_{yc}\}. \quad (37)$$

### B. Spectrum Alignment

To estimate and correct for the phase error, a basic assumption in most autofocus approaches is that the phase errors for different targets are space-invariant. The reason is twofold. First, if the phase error is space-invariant, the data will have much more redundancy, therefore can provide higher estimate accuracy for the phase error. Secondly, the space-invariant phase error can be corrected by a batch processing, which makes the correction process much more efficient.

However, the phase error spectrums for different targets in the FBP image have the same shape but different support areas. Therefore, the phase errors are space-variant. To facilitate the estimation and correction of the phase error, some preprocessing operations to align the phase error spectrums are required.

From (34), we have known that different targets have different spectrum displacements. Approximately, this spectrum displacement lies only in the azimuth dimension, and its size is linearly related to the azimuth position of the target in the spatial domain, i.e.,

$$\Delta k_x = \frac{k_y}{y_a(0)} x . \qquad (38)$$

To make the phase error spectrums coincide, alignment of the signal support in the spectrum domain can be achieved by a proper phase adjustment in the spatial domain. To take into account the $k_y$ dependence of the $k_x$ domain displacement, the phase correction should be performed in the $(x, k_y)$ domain. If we assume that the phase correction function is $\varphi_{cor2}(x, k_y)$, then it should satisfy the following condition

$$\frac{d\varphi_{cor2}(x, k_y)}{dx} = \Delta k_x . \qquad (39)$$

Inserting (38) into (39), it is easy to get the phase correction function as

$$\varphi_{cor2} = \frac{k_y}{2 y_a(0)} x^2 . \qquad (40)$$

*C. Azimuth Phase Error Estimate*

After the above two preprocessing operations, the spectrums of all targets in the illuminated scene are unambiguous and coincident. Therefore, the residual 2-D phase error can be estimated and corrected by a batch processing. In this approach, we incorporate the derived prior information on phase error structure, so only a 1-D phase error is required to estimate directly. The 1-D phase error can be either azimuth phase error or residual range cell migration. However, it is natural to choose the azimuth phase error because a variety of autofocus techniques to estimate the azimuth phase error are available in the literature.

The APE estimation can be implemented using a conventional 1-D autofocus algorithm. But some necessary modifications will be required. Firstly, the APE estimate may be affected by the residual RCM. The APE estimate is implemented in the range compressed data, and a customary presumption for most 1-D autofocus algorithms is that the scatterer's energy remains in a single range resolution cell. This requirement can't be met in 2-D defocus case. To solve this problem, a straightforward way is to perform a preprocessing on the data to reduce the range resolution, thereby keeping the residual RCM smaller than a coarse range resolution cell. After this preprocessing, the APE can be estimated directly by conventional autofocus techniques such as PGA. Secondly, an accurate APE estimate from a serious defocused image is still a challenging problem. When the APE across the whole aperture is very large or of high frequency, the imagery will be seriously defocused. Therefore, it is very difficult to extract some strong scatterers to estimate the APE. To solve this problem, we can use a divide and conquer strategy. First, the whole aperture is divided into several small subapertures [39]. As long as the length of each subaperture is small enough, the APE in each subaperture will become small enough then traditional autofocus methods can be used to extract the subaperture phase error. Finally, phase errors from all subapertures are then coherently combined to estimate the overall APE.

*D. 2-D Phase Error Computation and Correction*

Once the APE is estimated, instead of directly estimating the 2-D phase error, the proposed approach maps the estimated 1-D APE into the 2-D phase error by exploiting the analytical relationship between the 2-D phase error and the APE shown in (36).

Without loss of generality, we assume that the estimated APE is denoted as $\hat{\phi}_0(k_x)$, then the 2-D phase error can be calculated from the estimated $\hat{\phi}_0(k_x)$ directly by

$$\hat{\Phi}_e(k_x, k_y) = \frac{k_y}{k_{yc}} \hat{\phi}_0\left(\frac{k_{yc}}{k_y} k_x\right) . \qquad (41)$$

This mapping includes two steps. First, a range-frequency-dependent scaling transform is performed on the APE estimate. This scaling transform can be implemented by either interpolation, or chirp scaling techniques [40] . Then, the scaled APE estimate is multiplied by a range-frequency-dependent phase factor to obtain a 2-D phase error estimate.

Finally, the 2-D phase error is corrected for in the image spectrum domain using its estimate from (41). And then the corrected spectrum data is returned to the image domain to get a refocused imagery.

*E. Iteration*

In the proposed approach, the 2-D phase error estimate is computed from the estimated APE, therefore the accuracy of 2-D autofocus correction is completely determined by the accuracy of the APE estimate. However, accurate measurement of APE is limited by the residual RCM because the error energy is spread across several range resolution cells. Although some schemes, e.g., reduction of range resolution, or subaperture-based APE estimate, can be used to attenuate this effect, there still exist the cases where residual interaction can't be ignored. In this situation, it may be necessary to execute the estimation and correction process in an iterative manner. That is, after the image is corrected using the initial estimation of 2-D phase error, the entire process is repeated on this refocused image. Our experience has shown that 1-2 iterations will provide satisfactory results.

V. EXPERIMENTAL RESULTS

To verify the theoretical analysis and evaluate the performance of the proposed autofocus approach, both simulation and real data are processed and analyzed.

*A. Simulation Results*

Firstly, simulation experiments are performed to show the spectrum property of FBP imagery. The simulated SAR system operates in spotlight mode. Its system and imaging geometrical parameters are shown in Table I. Without losing of generality, we assume that the radar undergoes a nonlinear flight trajectory during the data collection process. The geometrical relationship between the radar and the illuminated scene is shown in Fig.8. In the illuminated scene, three point targets located at different positions are assumed.

TABLE I. SIMULATION PARAMETERS

| Parameter | Value |
| --- | --- |
| Carrier frequency | 10GHz |
| Range resolution | 0.12m |
| Azimuth resolution | 0.12m |
| Range | 15km |
| Radar altitude | 5000m |
| Nominal radar velocity | 120m/s |

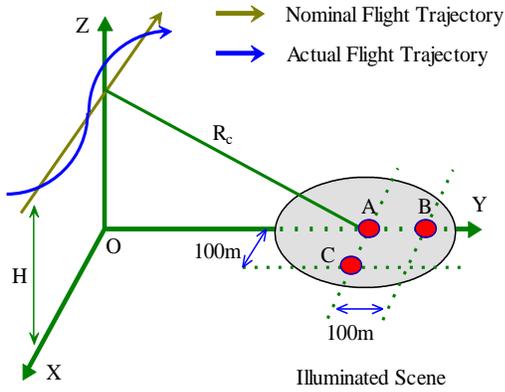

Fig. 8. Data collection geometry.

The simulated raw data is processed by the FBP. During the image formation processing, the nominal linear flight trajectory is assumed. Fig.9 shows the processing results, where Fig.9(a) is the FBP imagery and Fig.9(b) is the corresponding range-doppler domain imagery (returning the FBP imagery to azimuth frequency domain by an azimuth FFT). From both figures, it is obvious that the FBP imagery still suffers from serious 2-D defocus due to the uncompensated phase error resulted from the motion deviations.

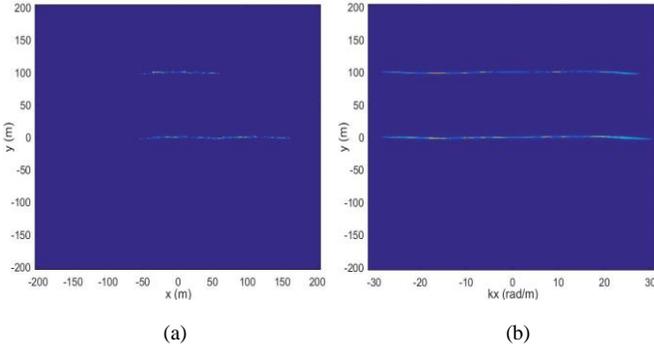

Fig.9. Processing results by FBP. (a) Full compressed imagery. (b) Range compressed imagery.

To analyze the 2-D imagery spectrum property, a 2-D FFT is performed on the above formed FBP imagery. Fig.10 (a) shows the resulted 2-D amplitude spectrum. To show more clearly the space-variant property of the imagery spectrum, the imagery spectrums of the separated target (target A, B and C) are also presented in Fig.10 (b), Fig.10(c), and Fig.10(d), respectively. As we all know, for PFA, there is no ambiguity in the spectrum domain, and the imagery spectrums for different targets possess the same spectrum support; for RMA, there is also no ambiguity in the spectrum domain, but the spectrum supports are different for different targets. However, for FBP, the imagery spectrum is observed in Fig.10 as aliased into baseband in the range wavenumber direction. Also, the spectrum supports are different for different scatterers. Specifically, the scatterers in different range positions share approximately the same spectrum support, but scatterers in different azimuth positions possess different spectrum supports.

To eliminate the spectrum ambiguity and align the spectrum supports for different scatterers, a preprocessing proposed in our 2-D autofocus approach is performed on the FBP imagery. After this preprocessing, the FBP imagery spectrum is shown in Fig.11. It is evident (obvious) that the spectrum ambiguity is eliminated and the spectrum supports of different scatterers are all aligned. Now the FBP imagery spectrum is almost the same as the PFA imagery spectrum. Therefore, efficient autofocus can now be applied to this preprocessed FBP imagery to estimate and correct for the residual 2-D phase error by a batch processing.

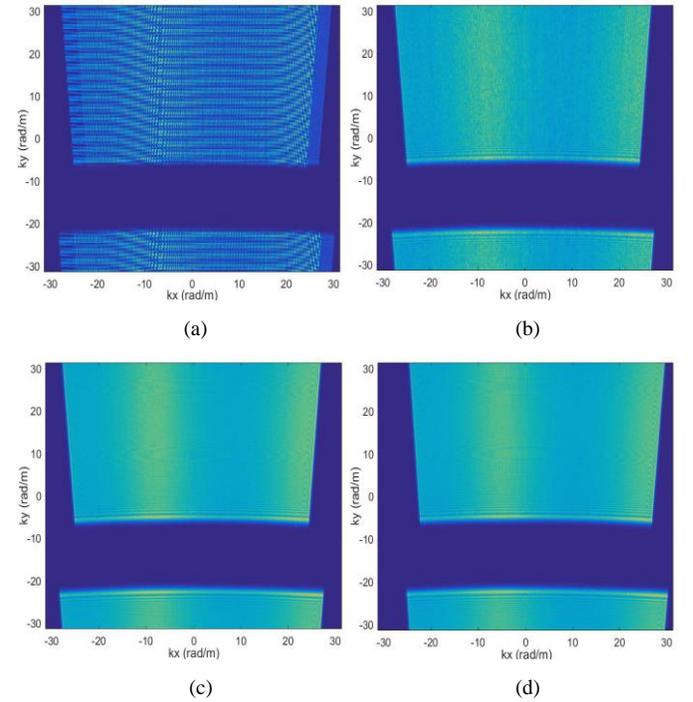

Fig.10. 2-D amplitude spectrum before preprocessing. (a) Target A+B+C. (b) Target A. (c) Target B. (d) Target C.

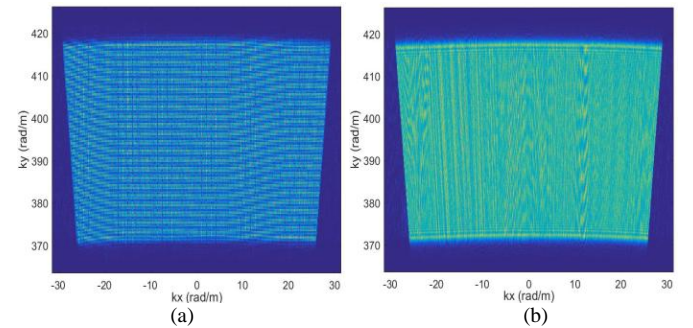

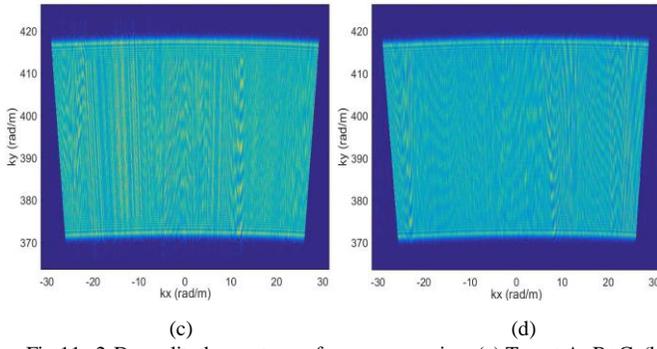

(c)                   (d)

Fig.11. 2-D amplitude spectrum after preprocessing. (a) Target A+B+C. (b) Target A. (c) Target B. (d) Target C.

To show the structure property of the residual 2-D phase error, the 2-D phase spectrum of target A is also presented. Fig.12 shows the 2-D phase spectrum and Fig. 13 shows its azimuth profiles evaluated at four different range spatial frequencies. To verify whether this measured 2-D phase spectrum satisfies the theoretical prediction shown in (36), the azimuth profiles at the same four range spatial frequencies are also directly computed from the azimuth profile of the 2-D phase spectrum at the center range frequency by exploiting the relationship shown in (36). The computed results are shown in Fig.14, and the difference between these theoretical predictions and the measured values are shown in Fig.15. It is clear that the measured 2-D phase spectrum satisfies the analytical relationship shown in (36) within the range of measured error.

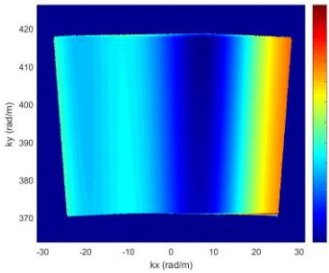 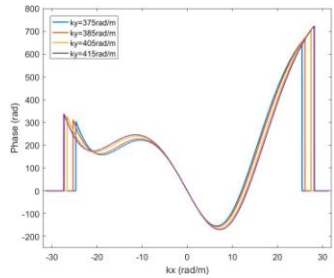

Fig.12. 2-D Phase spectrum of Target A.      Fig.13. Measured azimuth profiles of phase spectrum from Fig.12.

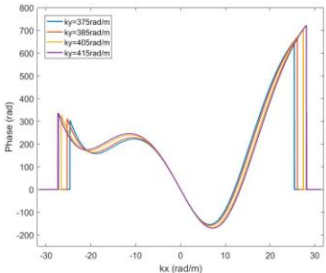 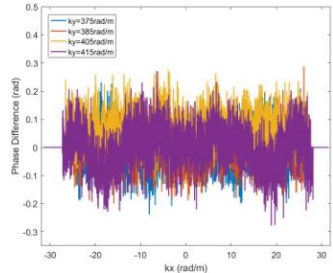

Fig.14. Computed azimuth profiles of phase spectrum by equation (36).     Fig.15. The difference between the measured and computed azimuth phase profiles.

Finally, the proposed 1-D estimation / 2-D correction autofocus approach is applied on the preprocessed FBP imagery, the refocused imagery is shown in Fig.16. To show more clearly the focus quality, the magnified point target responses of the three point targets are shown in Fig.17. From these figures, we can see that all the targets are well focused after the proposed 2-D autofocus processing.

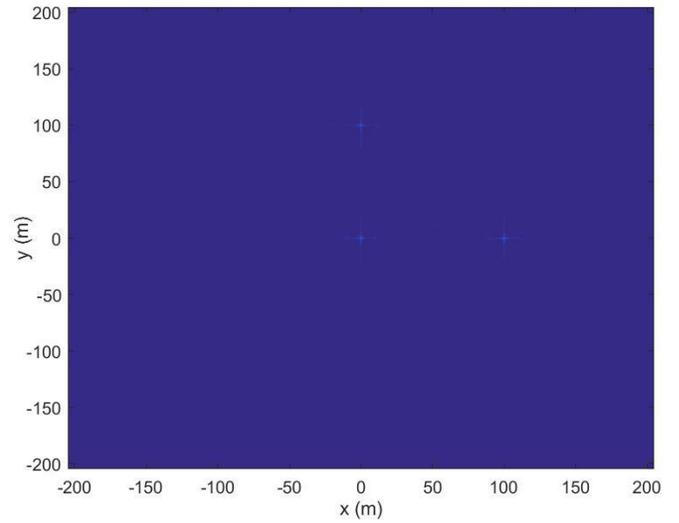

Fig.16. Refocused FBP imagery by the proposed 2-D autofocus approach.

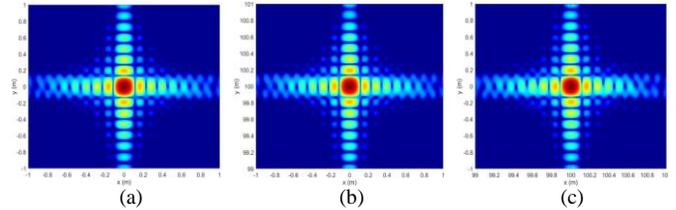

(a)            (b)            (c)

Fig.17. Magnified point target responses of the three targets in Fig.16. (a) Target A. (b) Target B. (c) Target C.

### B. Real Data Results

To verify the effectiveness of the proposed 2-D autofocus approach, data analysis and processing on a real SAR raw data are performed. The used raw data is collected by an ultra-high resolution airborne synthetic aperture radar operated in spotlight mode. The main parameters of this experimental radar are shown in Table II.

TABLE II. THE MAIN PARAMETERS OF THE RADAR

| Parameter | Value |
| --- | --- |
| Carrier frequency | 9.8GHz |
| Range resolution | 0.05m |
| Azimuth resolution | 0.04m |
| Range | 15km |
| Radar altitude | 6500m |
| Nominal radar velocity | 120m/s |

This radar has extreme-high spatial resolution in both range and azimuth direction. Therefore to produce an accurately focused imagery, a motion sensor with ultra-high accuracy is required to provide the geometric information which is necessary for the image formation processing. Although a luxury GPS/IMU sensor is equipped on the radar platform, its recorded position data is still not accurate enough to produce a completely focused imagery. Fig.18 shows the image produced by FBP processing on the raw data using the motion sensors information. It is clear that the imagery still suffers from serious 2-D defocus.

To show more clearly the 2-D defocus effect, a local area in Fig.19 and its corresponding range compressed image (return the FBP imagery to azimuth spatial frequency by an azimuth FFT) are enlarged and shown in Fig.20. From these figures, it is clear that the residual range migration exceeds several range resolution cells, therefore cannot be ignored in the followed autofocus processing.

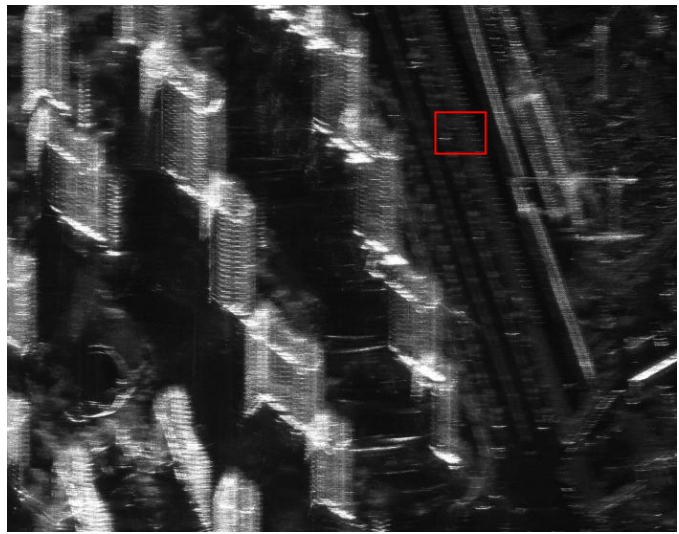

Fig.18. Imagery produced by FBP processing.

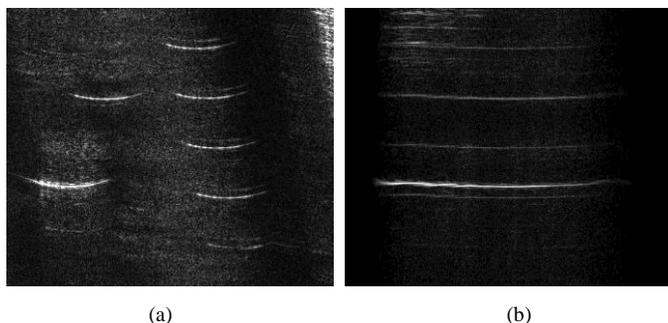

(a) (b)
Fig.19. Enlarged local area imagery in Fig.18. (a) Full compressed imagery. (b) Range compressed imagery.

To refocus the seriously defocused FBP imagery, the residual 2-D phase error resulted from the inaccurate motion measurements in the FBP imagery have to be estimated and compensated. To efficiently estimate and correct for the phase error by a batch processing, in our approach, a preprocessing on the FBP imagery is performed first to eliminate the spectrum ambiguity and align the spectrum support. Fig.20 (a) shows the 2-D amplitude spectrum of the FBP imagery. From the figure, we can see that the spectrum is observed as aliased into baseband in the range frequency domain and skewed in the azimuth frequency domain. After the first step preprocessing, as shown in Fig.20 (b), the observed spectrum is shifted to the center of the spectrum support in range direction therefore the ambiguity is eliminated. Then after the deskew preprocessing, the signal support areas for all scatterers are aligned, as shown in Fig.20 (c). The common phase error can then be extracted and corrected by using autofocus algorithms.

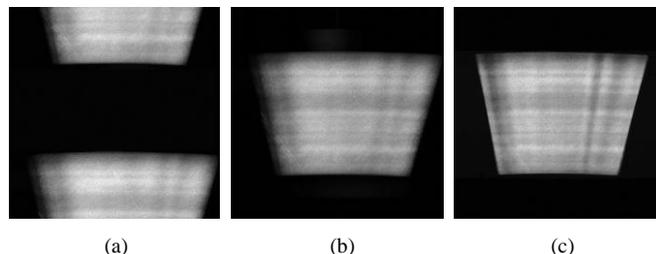

(a) (b) (c)
Fig.20. 2-D amplitude spectrum of the FBP imagery. (a) Before preprocessing. (b) After ambiguity elimination. (c) After spectrum alignment.

First, a 1-D autofocus processing (MD-PGA [39] is used in this experiment) is applied on the preprocessed FBP imagery and the result is shown in Fig.21. To show more clearly the focus quality, a local image of Fig.21 is enlarged and shown in Fig.22. Compared with the original FBP imagery, we can see that the focus quality of the refocused imagery after 1-D autofocus processing has been greatly improved. Nevertheless, a close look at the enlarged local imagery shows that the produced imagery after 1-D autofocus processing still suffers from defocus.

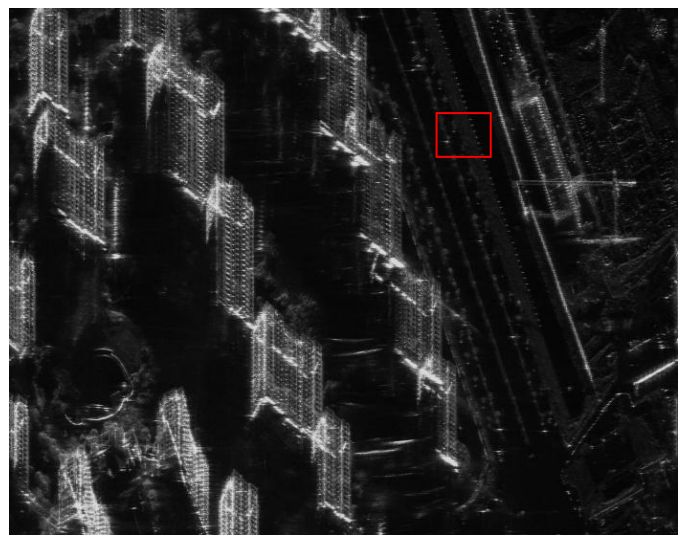

Fig.21. Refocused imagery after 1-D autofocus processing.

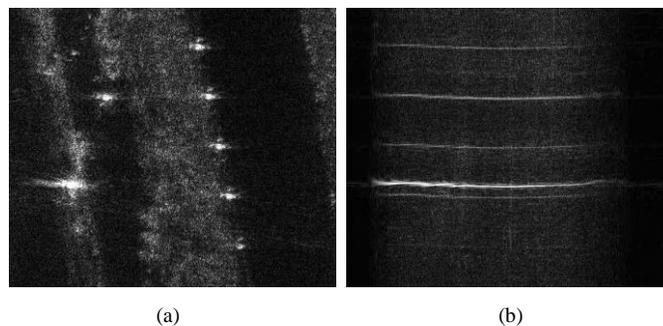

(a) (b)
Fig.22. Enlarged local area imagery in Fig.21. (a) Full compressed imagery. (b) Range compressed imagery.

Finally, the knowledge-aided 2-D autofocus method proposed in this work is applied on the preprocessed FBP imagery. In our approach, the APE is firstly estimated by the MD-PGA. Then, the residual 2-D phase error is directly computed from the estimated APE by using the analytical relationship shown in (36). Finally, the estimated 2-D phase error is eliminated from the defocused imagery. Fig.23 shows the produced refocused imagery. To see more clearly the improvement on the focus quality, a magnified local scene and its corresponding range compressed imagery are also shown in Figs.24. From these figures, we can clearly see that the 2-D degradation effects caused by the residual RCM are completely eliminated, and the produced imagery is well refocused.

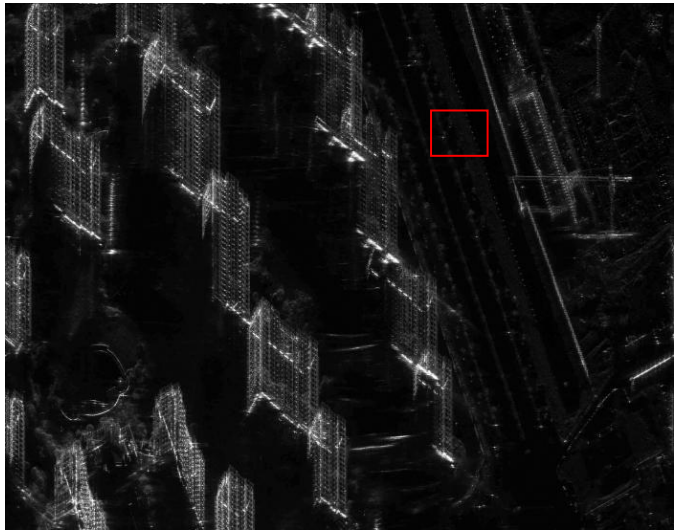

Fig.23. Refocused imagery after the proposed 2-D autofocus processing.

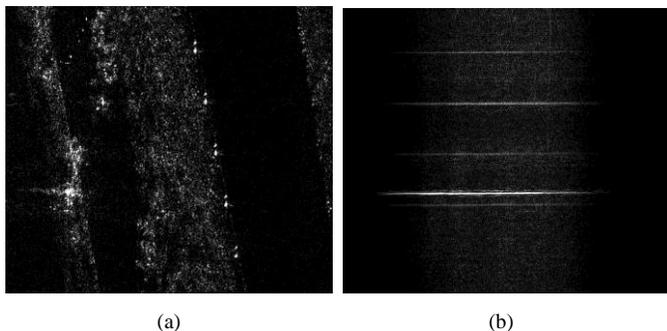

(a)                          (b)
Fig.24. Enlarged local area imagery in Fig.23. (a) Full compressed imagery. (b) Range compressed imagery.

To show the improvement of the focus quality quantitatively, two image quality indexes, image contrast and image entropy, are measured and shown in Table III. Both image indexes indicate a significant improvement in focus quality for the proposed approach.

TABLE III. COMPARISON OF IMAGE QUALITY INDEX

|  | FBP [8] (Fig.18) | 1-D autofocus [39] (Fig.21) | The proposed approach (Fig.23) |
|---|---|---|---|
| Contrast | 10.554 | 20.338 | 38.309 |
| Entropy | 16.736 | 16.138 | 15.584 |

## VI. CONCLUSION

In this paper, a new formulation of the filtered backprojection algorithm for spotlight synthetic aperture radar imaging was presented from the viewpoint of Fourier transform. This new interpretation shows clearly the analytical Fourier transform relationship between the phase history domain and the FBP imagery domain. By using this new formulation, the spectral characteristics of the FBP imagery, including the spectrum ambiguity, spatial-variant and structural property of the 2-D phase error in spectrum domain, are analyzed in detail. Then, by incorporating the *a priori* information on the property of the residual 2-D phase error, an accurate and efficient one-dimensional estimation/ two-dimensional correction 2-D autofocus approach is proposed. Because the phase error estimation is implemented in the dimension-reduced parameter space by a batch processing, the proposed strategy can significantly improve the autofocus performance in both focus accuracy and computational efficiency. Both simulation and real data processing results have verified the correctness of the theoretical analysis and the effectiveness of the proposed 2-D autofocus approach.

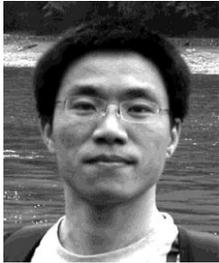

**Xinhua Mao** was born in Lianyuan, China, in 1979. He received the B.S. and Ph.D. degrees from the Nanjing University of Aeronautics and Astronautics (NUAA), Nanjing, China, in 2003 and 2009, respectively, all in electronic engineering.

Since 2009, he joined the Department of Electronic Engineering, NUAA, where he is now an associate professor. His research interests include radar imaging, and ground moving target indication (GMTI). He has developed algorithms for several operational airborne SAR systems and received the national defense technology award three times.